# Control and characterization of the preferential crystalline orientation of MoS$_2$ 2D flakes in printed layers

Camille Moisan[1], Zahra Gholamvand[1], Gabriel Monge[2], Karim Inal[2], Gilles Dennler[1]

[1] IMRA Europe SAS, 220 rue Albert Caquot, 06904 Sophia Antipolis Cedex, France.

[2] Mines ParisTech., Centre de Mise en Forme des Matériaux, UMR (Unité Mixte de Recherche) 7635, 1 rue Claude Daunesse, 06904 Sophia Antipolis Cedex, France.

*Corresponding authors: moisan@imra-europe.com; dennler@imra-europe.com

**ABSTRACT**

The recent development of Liquid Phase Exfoliation (LPE) of 2D materials has enabled the formulation of inks with rheological properties adapted to numerous liquid deposition methods. This has allowed the fabrication of various types of printed devices with unique features stemming from the nano-structure of the printed 2D layers. In this short communication, we demonstrate that the preferred crystalline orientation of printed MoS$_2$ flakes depends drastically upon the printing method employed to deposit the layers. Using angle resolved X-Ray Diffraction (XRD) to measure Pole Figure and subsequently calculate Orientation Distribution Functions (ODF), we show that the spin-coating method yields the best basal fiber texture, most likely because of the shear force at work on the flakes during the deposition process. This interim report thereby paves the way to further investigations and fine control of the preferred crystalline orientation of printed 2D flakes for the development of dedicated devices.

The recent development of Liquid Phase Exfoliation (LPE) of 2D materials[1] has enabled the formulation of inks with rheological properties adapted to numerous liquid deposition methods[2,3]. This allowed the fabrication of various types of devices like sensors, logic memory devices[4], and transistors[5]. It is notorious that both the size and the thickness of the flakes forming the printed layers have some tremendous effect on their physical and chemical properties[6]. However, to the best of our knowledge, the impact of the orientation of the flakes within the printed layers has not been investigated thoroughly yet. Moreover, only little is known to date about the effect of the printing method, the nature of the solvents and/or surfactant (Hansen solubility parameters, viscosity, boiling point) used, and the substrate to be coated (surface energy, temperature, roughness) upon the said crystallographic orientation of the flakes after printing. As a matter of fact, the precise characterization and quantification of the orientation is not straightforward. Indeed, while several papers report different technics for the study of single 2D materials flakes like Raman[7] or Scanning Tunneling Microscopy[8], the fact that printed layers comprise a large number of such flakes makes their precise characterization rather tedious. One notorious way to study the crystalline orientation of thin organic or inorganic films is the Grazing-Incidence Wide-Angle X-ray Scattering (GIWAXS)[9]. However, GIWAXS usually requires synchrotron beam lines, what hinders a systematic and easy usage thereof[10]. We report herein the employment of a standard X-Ray Diffraction (XRD) goniometer to characterize precisely the crystallographic orientation of $MoS_2$ flakes in printed layers. We show that through the measurement of a set of Pole Figures and the subsequent calculation of Orientation Distribution Functions (ODF)[11], this approach allows the assessment of the texture of the layer as well as a genuine quantitative comparison of the crystallographic orientation achieved by various printing methods.

Figure 1 show the diffractograms obtained by standard $2\theta$ XRD measurements on various types of $MoS_2$ samples. The so called "powder sample" displays all the peaks indicated in the $MoS_2$ 03-065-0160 (molybdenite-2H syn) ICSD datasheet (P63/mmc, 194) which has an hexagonal symmetry. The fact that the relative intensity of the peaks seems to follow qualitatively the ones of the ICSD datasheet indicates that the materials is unlikely to be textured, and that no preferential crystalline orientation seem to exist in this very sample.

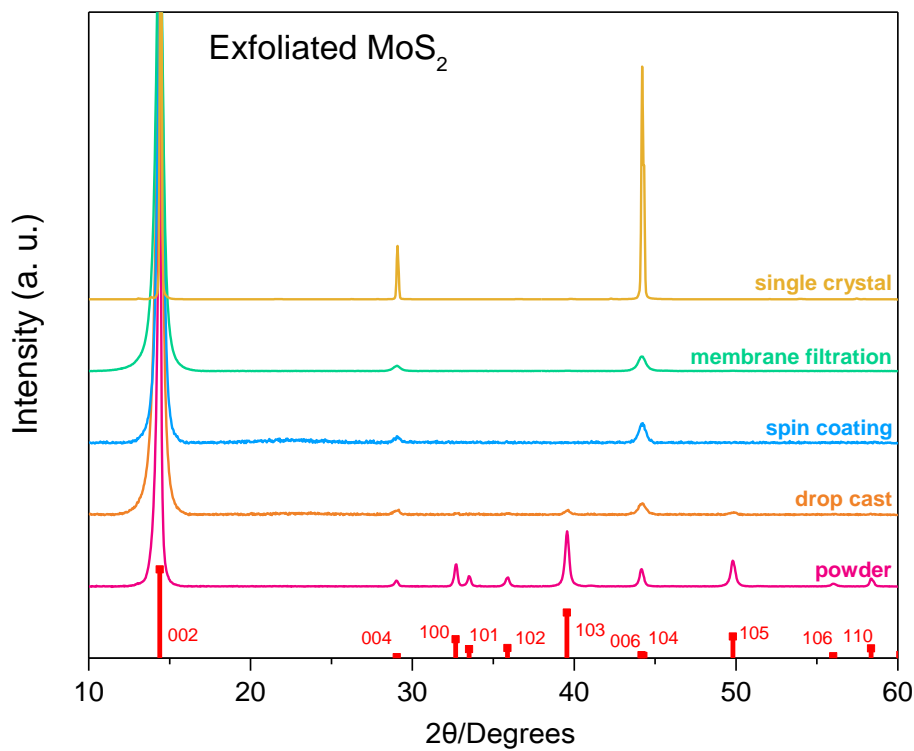

*Figure 1: Standard 2θ XRD diffractograms measured on various types of $MoS_2$ samples (single crystal, powder, layer stamped on glass from membrane filtration, layer drop casted on glass, and layer spin-coated on glass).*

The case of the single crystal is radically different: Only the {002}, {004} and {006} peaks are present. We can thereby conclude that this single crystal is oriented with the <00l> direction perpendicular to the substrate holder, and that the vast majority of the $MoS_2$ sheets comprised in this large crystal lay parallel to the substrate holder along their crystallographic basal plane. The respective diffractograms of the membrane filtration, the spin-coated and the drop casted samples look all quite similar (see METHOD section for details on the sample preparation). The latter one displays some tiny signal for the {103} and {105} peaks, what suggests a slightly more disordered structure. But all three layers seem comprised mostly of $MoS_2$ flakes laying parallel to the substrate and the sample holder. Thus in spite of the fact that this measurement technique allows one to conclude that all printed/deposited layers of $MoS_2$ are most likely textured with a preferential orientation of the 2D sheet parallel to the substrate, it prohibits any precise and quantitative comparison. This approach is thereby not much more informative than standard Scanning Electron Microscope (SEM) top views (Figure S1 of the Supplementary Information).

Figure 2 illustrates the power of genuine texture investigations carried with a XRD setup equipped with a four circle goniometer. Figure 2a, 2b, 2c and 2d represent the {103} Pole Figures of the $MoS_2$ single crystal, the $MoS_2$ powder, the $MoS_2$ layer stamped on glass from membrane filtration, and the $MoS_2$ solution spin-coated on glass, respectively. While Figure 2a is comprised of 6 distinct and well defined spots corresponding to the multiplicity of the {103} family planes of the hexagonal symmetry of the $MoS_2$ crystal, Figure 2c and 2d are basically made of one single tore: This indicates that for these two printed samples, the flakes have a tendency to lay rather parallel to the substrate, yet with no preferred orientation in-plane. These features indicate a so-called "fiber texture". The Pole Figures provide some precious additional information regarding the degree of orientation: For example, the tore described by the Pole Figure of the spin-coated sample (Figure 2d) is noticeably thinner than the one recorded in the case of the membrane filtered sample (Figure 2c). This proves without ambiguity that the latter sample is less organized than the former one. As for the Figure 2b, no clear preferred orientation appears in this Pole Figure. The tores observed in Figure 2c and 2d are here very wide and barely formed. Interestingly though, the {002} Pole Figure of this very sample (Figure S2 of the Supplementary Information) suggests the presence of a weak texture, completely invisible in Figure 1, and further confirmed by the calculation of the Orientation Distribution Function (see below). We believe that this unexpected light anisotropy was induced by the sample preparation (see METHOD).

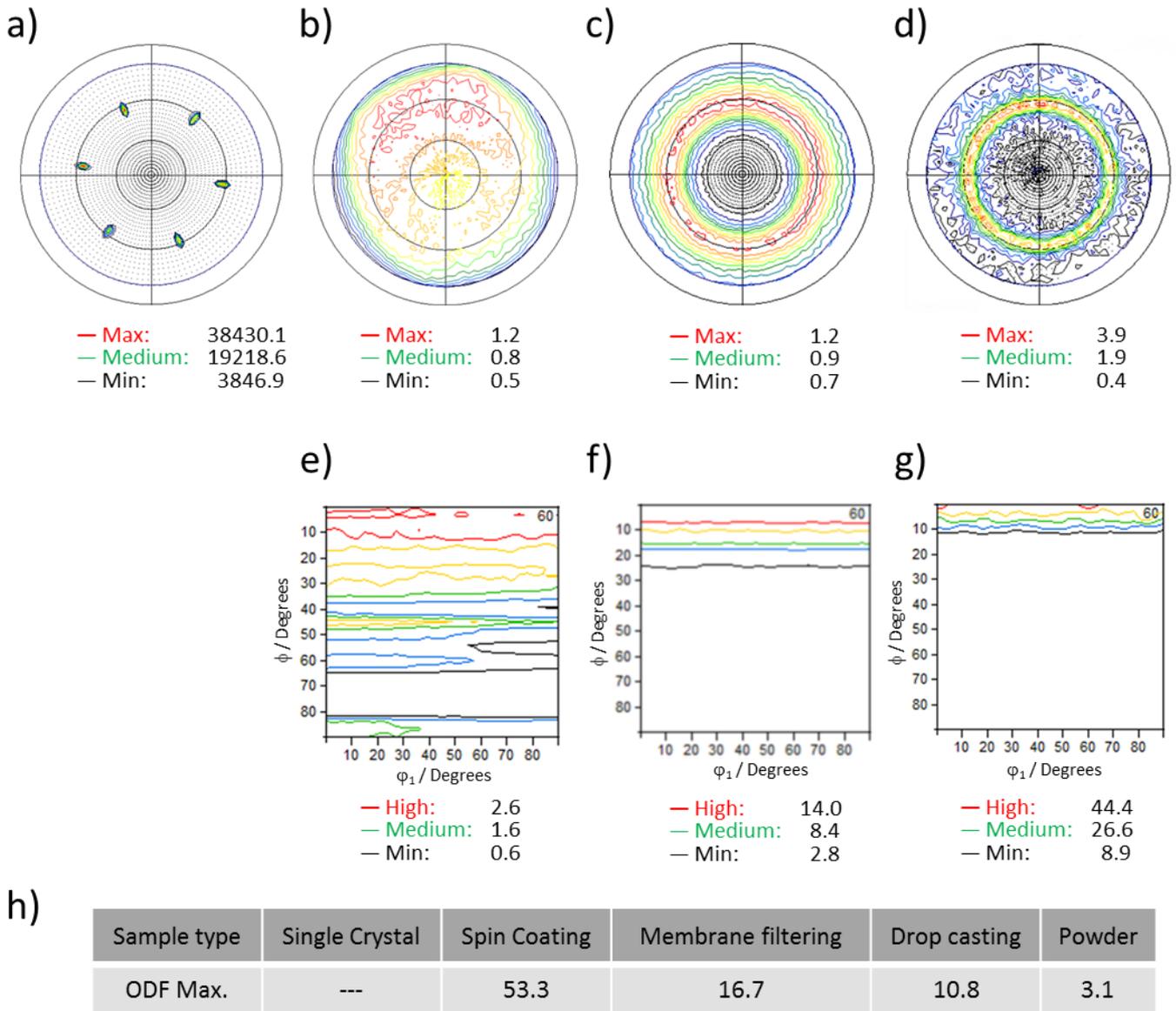

Figure 2: {103} Pole Figures of a) the MoS$_2$ single crystal, b) the MoS$_2$ powder, c) the MoS$_2$ layer stamped on glass from membrane filtration, and d) the MoS$_2$ solution spin-coated on glass; Orientation Distribution Functions (ODF) represented in the ($\varphi_1, \varphi$) plan for the section $\varphi_2=60°$ for e) the MoS$_2$ powder, f) ) the MoS$_2$ layer stamped on glass from membrane filtration, g) the MoS$_2$ solution spin-coated on glass; h) Table showing the maximum of the ODF for each type of sample studied.

Measuring 3 independent Pole Figures allows one to calculate the so-called Orientation Distribution Function (ODF)[11]. The ODF is a quantitative measure of the probability of a given crystalline orientation in the Euler space ($\varphi_1, \varphi, \varphi_2$), expressed in Multiples of a Random Distribution (mrd). Note that a mrd of 1 for a given orientation means that this said orientation has the same probability to occur than in an entirely disorganized powder.

We have thus recorded the {002}, the {103} and the {105} Pole Figures of all our MoS$_2$ samples (Figure S3 of the Supplementary Information), and calculated the respective ODF therefrom. Figure 2e, 2f, and 2g display the ODF of the MoS$_2$ powder, the MoS$_2$ layer stamped on glass from membrane filtration, and the MoS$_2$ solution spin-coated on glass. All these ODFs are represented in the $(\varphi_1, \varphi)$ plan for the section $\varphi_2=60°$, and the color code mentioned below each graph indicates the maximum, the medium and the minimum values of these ODFs. Figure 2f and 2g feature a typical fiber texture displaying equal values of $\phi$ whatever the value of $\varphi_1$ and $\varphi_2$ are (see Figure S4 of the Supplementary Information). In the case of the spin-coated sample, the ODF reaches a maximum value of almost 45 mrd, and plummets for $\phi \geq 10°$. For the membrane filtered sample, the ODF hits a maximum of 14 rmd, and reduces significantly for $\phi \geq 20°$. The Table presented in Figure 2h summarizes the maximum ODF reached for the all the various samples. It appears clearly that the spin coating approach is the deposition method that yields the most-fiber textured layers. We speculate that this observation can be partially explained by the centrifuge force applied to the MoS$_2$ platelets during the spin coating step. This strong shear force may help the flakes to reach a position parallel to the substrate, and perpendicular to the axis of rotation of the spin-coater. We thereby believe that the spinning speed, the viscosity of the solvent as well as its boiling point, may have a drastic influence of the preferential crystalline orientation of the flakes.

As a summary, we could demonstrate in this short communication that angle resolved XRD is a very powerful tool for the quantification of the preferential crystalline orientation of 2D material flakes deposited as thin layers by various methodologies. We observed that the spin-coating method yield the best fiber texture, most likely because of the shear force at work on the flakes during the process. Now that this first conclusion has been drawn and that a method allowing a systematic characterization has been proposed, we intent to study the effect of the nature of the solvent (viscosity, boiling point, etc…), the spinning-speed, and the size of the 2D flakes (amongst other parameter) upon the ODF of the layer deposited, in order to hopefully reach values close to single crystals.

**METHODS**

The $MoS_2$ single crystal was purchased from HQ Graphene with the dimensions 0.5-0.7 cm. $MoS_2$ dispersions were prepared by a method described in a previous work[6]. $MoS_2$ powder was purchased from Sigma-Aldrich (<2 µm) and dispersed in a 70 mL aqueous surfactant (sodium cholate SC, $C_{cs}$ = 10 g/L) solution with an initial concentration of 20 g/L. The mixture was sonicated (1 h at 60% amplitude) with a high power sonic probe (VibraCell CVX; 750 W, 60 kHz) equipped with a flathead tip. The resulting dispersion was then centrifuged in 10 mL vials in Hettich Mikro 220R centrifuge equipped with a horizontal rotor 1020 at 5000 rpm for 90 min. The supernatant, was discarded. The sediment was re-dispersed in the solution of 1 g/L sodium cholate to a volume of 70 mL and subjected to the second sonication for 5 h at 60% amplitude, pulse rate 6 on 2 s off to exfoliate $MoS_2$ layered crystals. This gives a polydisperse $MoS_2$ stock dispersion containing both un-exfoliated crystallites and exfoliated $MoS_2$ nanosheets. The size selection procedure was adapted from a previously reported liquid cascade centrifugation[12].

As angle resolved XRD measurements require to tilt the sample up to a large angle (see below), we had to embed the powder in an inert matrix. This was performed by mixing $MoS_2$ powder with Epoxy glue. The mixture was then dropped on a flat glass substrate and let dry at room temperature. Then the dried deposit was removed from the substrate and the flat side was analyzed by XRD.

The membrane samples were obtained by vacuum filtering $MoS_2$ dispersions through porous mixed cellulose ester filter membranes (MF-Millipore membrane, hydrophilic, 0.025 µm pore size, diameter of 47 mm). To remove the remaining surfactant, all the films were washed by filtering 50 mL of deionized water through the porous films. The resulting films once dried, were cut to the desired dimensions and transferred onto glass. The cellulose membrane was removed by fixing the deposit film facing the substrate with IPA and applying pressure to the film, taking it off with acetone vapour, and subjecting it to a series of acetone baths to remove any traces. The acetone dissolves the cellulose membrane and leaves the porous films behind on the substrate surface.

The spin coated sample was made in a $N_2$ filled glove box with a spin coater (Suss Microtec DELTA6RC). 100 µL of the dispersion was poured on a glass substrate before spinning. Then the sample was subjected to a rotation for 20 second at 6000 rpm and let dry at room temperature in the glove box.

SEM images were acquired with a Hitachi S-4700 (field emission) apparatus. 2θ XRD was measured with a Bruker D8 Advanced diffractometer using Cu Kα radiation (parallel beam geometry).

The XRD texture measurements were performed using a Panalytical Empyrean diffractometer with Cu-Kα radiation (45 kV, 30 mA) in parallel beam configuration (poly-capillary X-ray lens for the incident beam and long parallel plate collimator for the diffracted beam). The textures were examined by measuring the three incomplete (to a tilt angle of 81°) pole figures {002}, {103}, and {105}. The Orientation Distribution Functions (ODFs) were computed from the above three pole figures using X'Pert Texture software of Panalytical, after background intensity subtraction and defocusing correction (even if this last effect is minimized using poly-capillary lens). Usually, ODF can be displayed as a three dimensional plot with the three Euler angles as axes. For hexagonal material, the ODF data are normally shown as a series of sections taken through the three dimensional ODF space at φ2=0, 5, 10…90°.

## DATA AVAILABILITY

All relevant data are available from the authors.

## ACKNOWLEDGEMENT

The authors would like to thank Prof. Pascal Roussel for fruitful scientific discussions on the topic of XRD diffractograms.

## CONTRIBUTIONS

CM prepared all the samples under the supervision and the technical guidance of ZG. Moreover, she performed the SEM experiments and the 2θ XRD measurements. GB and KI carried out all the Pole Figure measurements and extracted the ODF for all samples. GD supervised the work and produced the manuscript, with the help of all.

## ADDITIONAL INFORMATION

Supporting Information accompanies the paper on the npj 2D Materials and Applications website.

**Competing interests**: The authors declare that they have no competing financial interests.

# Control and characterization of the preferential crystalline orientation of MoS$_2$ 2D flakes in printed layers
# (Supporting Information)

Camille Moisan[1*], Zahra Gholamvand[1], Gabriel Monge[2], Karim Inal[2], Gilles Dennler[1*]

[1] IMRA Europe SAS, 220 rue Albert Caquot, 06904 Sophia Antipolis Cedex, France

[2] Mines ParisTech., Centre de Mise en Forme des Matériaux, UMR (Unité Mixte de Recherche) 7635, 1 rue Claude Daunesse, 06904 Sophia Antipolis Cedex, France

*Corresponding authors: moisan@imra-europe.com; dennler@imra-europe.com

**Scanning Electron Microcopy images**

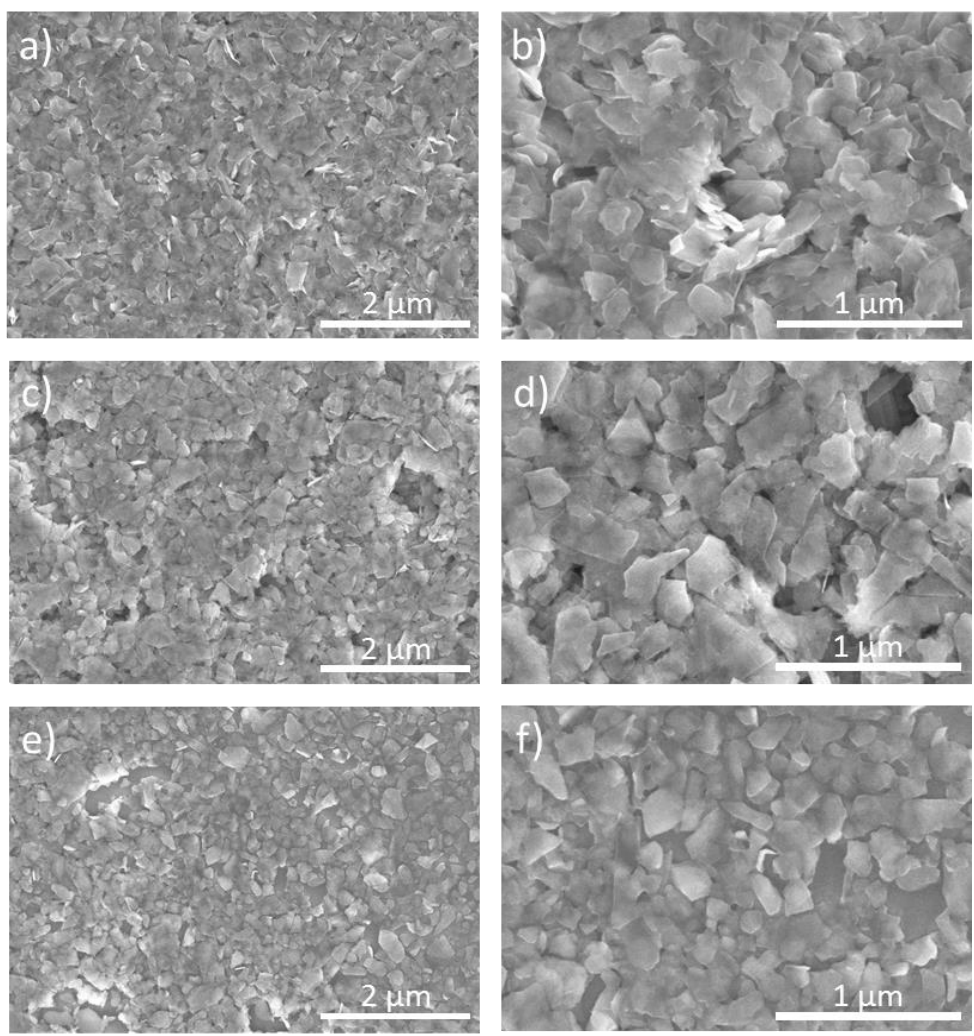

*Figure S1: SEM pictures (top views) of the drop casted sample (a and b), the membrane filtered sample (c and d), and the spin coated sample (e and f).*

Figure S1 represents SEM top views of the drop casted sample (a and b), the membrane filtered sample (c and d), and the spin coated sample (e and f). It shows that such images, although very well resolved, do not provide any objective information about the relative preferred crystalline orientation in each sample. Basically, all surfaces appear equally rough (or flat, depending of the view point), and no further conclusion can be drawn from them. It should be noted here that we tried to perform cross section views, but the layers were too thin to allow any high quality images.

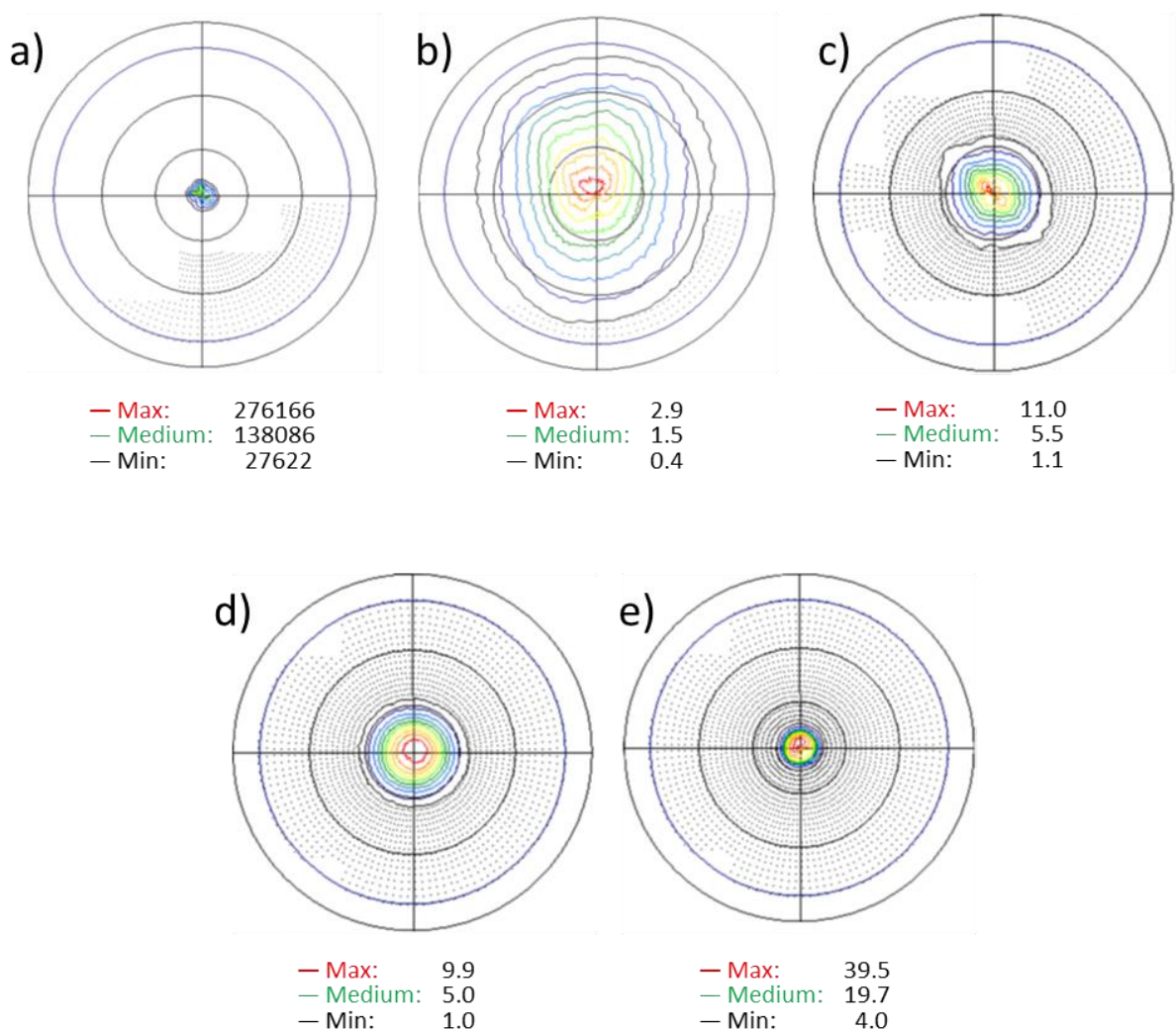

*Figure S2: {002} Pole Figures of a) the MoS$_2$ single crystal, b) the MoS$_2$ powder, c) the MoS$_2$ layer drop-casted on glass, d) the MoS$_2$ layer stamped on glass from membrane filtration, and e) the MoS$_2$ layer from solution spin-coated on glass.*

Figure S2 shows the {002} Pole Figures of all five different types of samples investigated. All these Pole Figures are comprised of circles centered on the middle point. However, the width and the intensity of these said circles depend drastically upon the nature of the samples. In the case of the single crystal, the XRD intensities measured are very high, and strongly concentrated toward the middle of the graph. This indicates that the sample is thick and highly crystalline (strong signal), highly oriented (circles having a small radius), yet not defect free (circles having a non-zero radius). The weaker the samples are fiber textured, the larger the circles, and the lower the intensity of the signal gets.

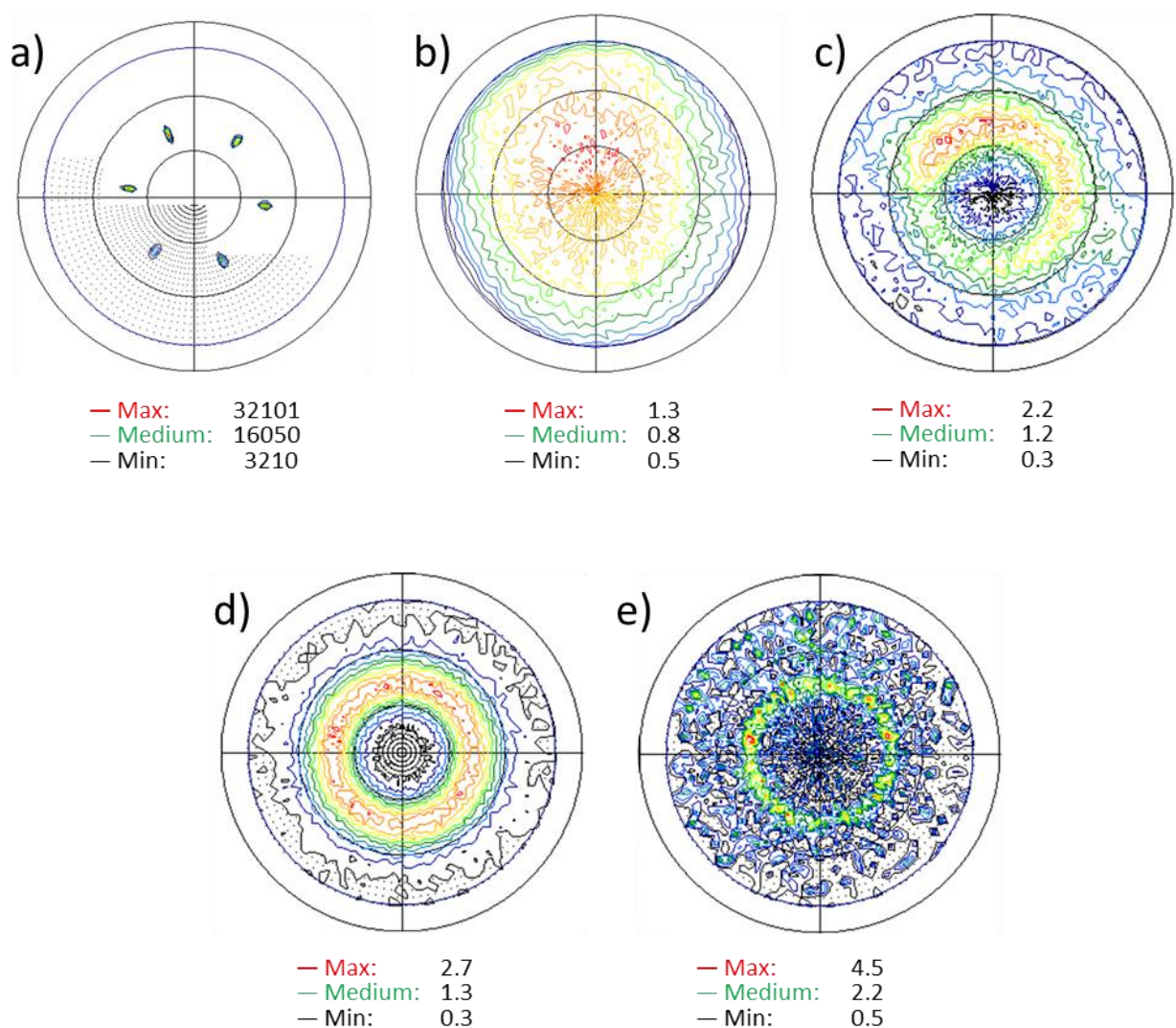

*Figure S3: {105} Pole Figures of a) the MoS$_2$ single crystal, b) the MoS$_2$ powder, c) the MoS$_2$ layer drop-casted on glass, d) the MoS$_2$ layer stamped on glass from membrane filtration, and e) the MoS$_2$ layer from solution spin-coated on glass.*

Figure S3 shows the {105} Pole Figures of all five different types of samples investigated. They resemble {103} Poles Figures shown in Figure 2, with the only difference that their signal appears at a lower tilt angle, due to the lowest angle between {002} and {105} in comparison with the angle between {002} and {103}.

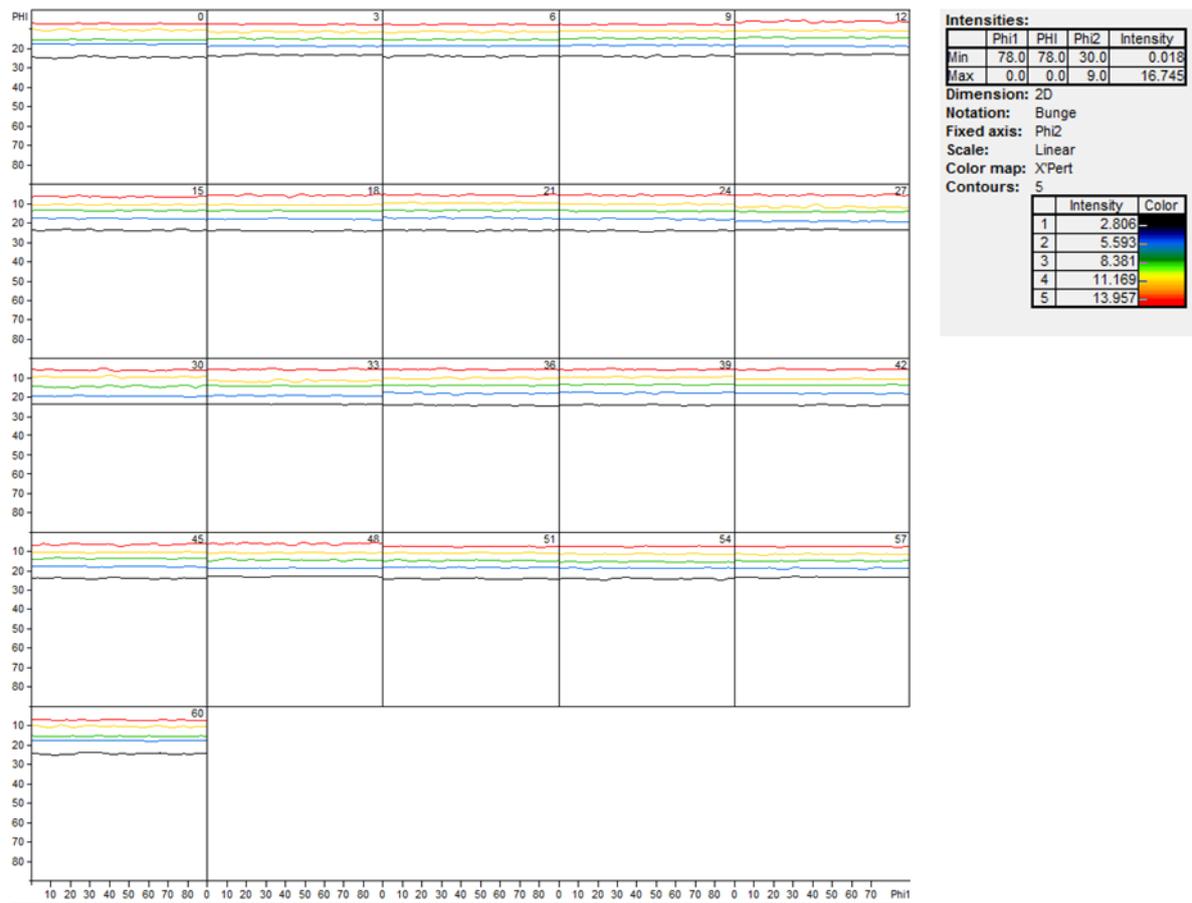

*Figure S4: Orientation distribution function (ODF) of the membrane filtered sample. The x axis of each square graph represents the $\varphi_1$ Euler angle, while the y axis represents $\phi$. Each square graph is a cross section of the ODF at a $\varphi_2$ Euler angle equal to the value mentioned in the upper right corner of the respective graphs.*

Figure S4 shows the Orientation distribution function of the membrane (ODF) filtered sample. The x axis of each square graph represents the $\varphi_1$ Euler angle, while the y axis represents $\phi$. Each square graph is a cross section of the ODF at a $\varphi_2$ Euler angle equal to the value mentioned in the upper right corner of the respective graphs. The isolines visible on this Figure are typical of a basal fiber texture.